\documentstyle[aps,preprint,prc]{revtex}
\date{\today}
\begin{document}
\flushbottom

\widetext
\draft
\title
 {\bf P,T-Violating Nuclear Matrix Elements in the
One-Meson Exchange Approximation}
\author
 {I.S. Towner and A.C. Hayes}
\address
{AECL Research, Chalk River
Laboratories, Chalk River, Ontario, Canada K0J 1J0}
\date{\today}
\maketitle

\def\thepage{\arabic{page}}
\makeatletter
\global\@specialpagefalse
\ifnum\c@page=1
\def\@oddhead{Draft\hfill To be submitted to Phys. Rev. C}
\else
\def\@oddhead{\hfill}
\fi
\let\@evenhead\@oddhead
\def\@oddfoot{\reset@font\rm\hfill \thepage \hfill}
\let\@evenfoot\@oddfoot
\makeatother

\begin{abstract}
Expressions for the P,T-violating NN potentials are derived for $\pi$,
 $\rho$ and
$\omega$ exchange.  The nuclear matrix elements for $\rho$ and
$\omega$ exchange are shown to be greatly suppressed, so that,
under the assumption of comparable coupling constants,
$\pi$ exchange would dominate by two orders of magnitude.
  The ratio
of P,T-violating to P-violating matrix elements is found to
 remain approximately constant across the nuclear mass
table, thus establishing the proportionality between time-reversal-violation
and parity-violation matrix elements. The calculated
values of this ratio suggest
a need to obtain an accuracy of order $ 5 \times 10^{-4}$ for the ratio
 of the PT-violating to P-violating asymmetries in
neutron transmission experiments in order to improve on the present
limits on the isovector pion coupling constant.
\\
{\bf PACS: {11.30.Er,13.75.Cs,21.30.+y,24.80.-x}}
\end{abstract}

\makeatletter
\global\@specialpagefalse
\def\@oddhead{\hfill}
\let\@evenhead\@oddhead
\makeatother
\nopagebreak
%\twocolumn
\onecolumn
\narrowtext

\section{INTRODUCTION} \label{sec:intro}
Simultaneous violation of parity conservation and time reversal invariance
(P,T-violation) in the low-energy nucleon-nucleon interaction can be described,
 in
analogy with the description of P-violation, in terms of nonrelativistic
potentials
derived from single-meson exchange diagrams involving the lightest pseudoscalar
 and
vector mesons \cite{one}.  We consider here $\pi$, $\rho$ and $\omega$ mesons.
  The
strength of the P,T-violation in the N-N interaction is then parameterised in
 terms of
the coupling constants, $\overline{g}^{(I)\prime}_{MNN}$, characterising the $N
\rightarrow NM$ matrix elements of the various isospin ($I$) components.  We
consider
the forms for the Lagrangian describing the $N \rightarrow NM$ vertex
 and find
in a calculation in nuclei described by a closed-shell-plus-one configuration
 that the
contribution to P,T-violation from all isospin components
of $\rho$-exchange are
identically zero for charge-symmetric ($N=Z$) closed-shell
cores. A similar result has been found \cite{one,eight} to
hold true for the $I=0$, $I=2$ components of
$\pi$-exchange. In
heavy nuclei with closed-shell cores that are not charge symmetric, assuming
equal couplings $\overline{g}^{(I)\prime}_{MNN}$, these
 terms together with the $\omega$-exchange terms remain
small compared to the dominant term coming from the $I=1$ component of
 $\pi$-exchange.
\par
P,T-violation may be studied in neutron transmission \cite{two}
 and gamma-decay \cite{three,four,five}
experiments on oriented nuclei.  In both cases P,T-violation is accompanied by
P-violation alone.  Consider a simple $\gamma$-decay example \cite{one,four}
 where the initial
nuclear state is of mixed parity due to the presence of P-violation
and P,T-violation
in the NN-interaction, while the final state to a good approximation remains
a state
of definite parity.  Let $|a>$ and $|b>$ denote the initial and final
 nuclear states
in the absence of P- and P,T-violation.  The potential, $V^P + V^{P,T}$,
has the
effect of changing $|a>$ to a state $|A>$ and $|b>$ to a state $|B>$,
 which under
these assumptions are given by
\begin{eqnarray}
|A>& \simeq & |a> + \sum_{a^\prime} {|a^\prime> \over E_a-E_{a^\prime}}
{}~\times
\nonumber
\\
&&~~~~~~~~~~ (<a^\prime|V^P|a> + <a^\prime|V^{P,T}|a>)
\end{eqnarray}
\begin{equation}
|B> \simeq |b>,
\end{equation}
\noindent where the state $|a^\prime>$ is of the same angular momentum as $|a>$
 but of
opposite parity.  The P,T-violating effect is then proportional to
\begin{equation}
A_{P,T} \simeq Im {<B\| \pi^\prime L^\prime \| A> \over
<B \| \pi L \| A>},
\end{equation}
\noindent where $\pi^\prime L^\prime$ is the irregular multipole now available
 due to
P,T-violation while $\pi L$ is the regular multipole present in the
 $|a> \rightarrow |b> + \gamma$ transition.
 A P-violating effect in the same transition is
 proportional
to
\begin{equation}
A_P \simeq Re
{<B\| \pi^\prime L^\prime \| A> \over
<B \| \pi L \| A>}.
\end{equation}
\noindent It is useful to consider the ratio of these effects, which becomes
proportional to the matrix elements of $V^P$ and $V^{P,T}$, viz \cite{one}
\begin{equation}
{A_{P,T} \over A_P} = -i {<a^\prime|V^{P,T}|a> \over
<a^\prime|V^P|a>} + \cdots
\end{equation}
\noindent Here the ellipsis indicates that effective T-violation contributions
induced by T-invar\-iant higher-order processes have been omitted.
 It is important
that these be estimated \cite{five} before any comparison is made between
 theory and
experiment, but they are not relevant for the discussion presented here.
 Herczeg \cite{one}
argues that if a calculation of $<a^\prime|V^{P,T}|a>$ is not available,
 a rough
estimate of the ratio $A_{P,T}/A_P$ combined with a measured value of
the P-violating
effect may give a better estimate of $<a^\prime|V^{P,T}|a>$ than a rough
 estimate of
$<a^\prime|V^{P,T}|a>$ itself.  To this end, Herczeg defines
\begin{equation}
-i {<a^\prime|V^{P,T}|a> \over <a^\prime |V^P|a>} =
\kappa^{(1)} {\overline{g}^{(1)\prime}_{\pi NN} \over g^{(0)\prime}
_{\rho NN}} .
\end{equation}
\noindent Here it is assumed that
P-violation occurs primarily through isoscalar ($I=0$) $\rho$-exchange
 characterized
by the coupling constant $g_{\rho NN}^{(0)\prime}$.  In the work
of Desplanques,
Donoghue and Holstein \cite{six} (DDH), an isovector ($I=1$) $\pi$-exchange
term is shown
potentially to be of comparable importance in P-violation.
However the limits on
measurements \cite{seven} of circular polarisation of a selected
 $\gamma$-transition in $^{18}F$,
which is sensitive only to the isovector component of the P-violation,
 is found to be
smaller than the DDH `best estimate' and at present there is only
 an upper limit.
 Thus, in this
analysis, we consider only isoscalar $\rho$-exchange contributing to
P-violation.
\par
Our goal here is to derive the potentials for the $\rho$- and $\omega$-
exchange P,T-violating NN potentials and to determine the strength of their
nuclear matrix elements relative to those for $\pi$- exchange.
%For this purpose we will always be assuming equal couplings
%$\overline{g}^{(I)\prime}_{MNN}$ for all isospin components of
%all three meson exchanges.
%We derive the  corresponding effective one-body potentials
We use these results to
 estimate a reasonable range for the value of
$\kappa^{(1)}$. One of our main
interests is in studying the dependence of the P,T-violating matrix elements,
and hence $\kappa^{(1)}$, on
nuclear mass.
 A somewhat
similar undertaking has be made by Griffiths and Vogel \cite{nine} for the
one-pion exchange potential. These authors follow Herczeg and express
 the effective
one-body interaction in terms of an asymptotic potential obtained
 assuming  $m_{\pi} \rightarrow \infty$.
 The true
one-body P,T-violating potential is then related to the asymptotic potential
through a suppression factor F, and F is shown to have a large dependence
on nuclear mass. In the present work we do
not use this analytic approximation as we find the assumption of
large meson mass not to be good, particularly in the case of the pion;
 rather we perform exact numerical
evaluations of the matrix elements.  We find
both the P,T- or P- violating matrix elements and consequently
$\kappa^{(1)}$
to be approximately mass independent.   Thus, we conclude that any
strong mass dependence obtained using asymptotic approximations
do not apply to the true potentials. In this regard there
is little advantage in choosing examples
in heavy nuclei over light as candidates
for experimental study.
Of course, any advantage from dynamical enhancements, such as the
smallness of the energy denominator in eq. (1), remains valid.

\section{THE P,T-VIOLATING INTERACTION} \label{sec:inter}
In a meson-exchange model, the P,T-violating nucleon-nucleon interaction
is given by
one weak P,T-violating $N \rightarrow NM$ vertex and one strong
 P,T-conserving $N \rightarrow NM$ vertex with a meson exchanged
 between the two.  For $\pi$-exchange the
weak vertices are described by Lagrangians:
$$
{\cal L}^{(I=0)}_{\pi NN} = \overline{g}^{(0)\prime}_{\pi NN}
\overline{N} N \tau^a \phi^a_{\pi},
$$
$$
{\cal L}^{(I=1)}_{\pi NN} = \overline{g}^{(1)\prime}_{\pi NN}
\overline{N} N \phi^z_{\pi},
$$
\begin{equation}
{\cal L}^{(I=2)}_{\pi NN} = \overline{g}^{(2)\prime}_{\pi NN}
\overline{N} N (3 \tau^z \phi^z_{\pi} - \tau^a \phi^a_{\pi}),
\end{equation}
where $\overline{g}^{(I)\prime}_{\pi NN}$ are coupling constants, $N$
 nucleon fields
and $\phi_{\pi}$ pion fields.  The Roman superscripts are Cartesian isospin
 indices
and a repeated index, $a$, is summed over, but not the index, $z$.  The strong
interaction vertex is
\begin{equation}
{\cal L}_{\pi NN} = i g_{\pi NN} \overline{N} \gamma_5 N \tau^a \phi^a_{\pi},
\end{equation}
\noindent with $g_{\pi NN}$ the strong pion-nucleon coupling constant,
and $\tau$ the
Pauli isospin matrix.  Combining one strong and one weak vertex Lagrangian
 with the
pion propagator, taking a nonrelativistic limit of the vertex functions, and
 making a
Fourier transform to coordinate space leads to the following expressions for a
P,T-violating potential
$$
V^{(I=0)}_{P,T} = \overline{g}^{(0)\prime}_{\pi NN} g_{\pi NN}
{m^2_{\pi} \over 8 \pi M}
 \mbox{\boldmath $\sigma_- \cdot$}  \hat{\mbox{\boldmath $r$}}
\mbox{\boldmath $\tau_1 \cdot \tau_2$} Y_1(x_{\pi}),
$$
$$
V^{(I=1)}_{P,T} = \overline{g}^{(1)\prime}_{\pi NN} g_{\pi NN}
{m^2_{\pi} \over 16 \pi M}
 [\mbox{\boldmath $\sigma_+ \cdot$} \hat{\mbox{\boldmath $r$}} \tau_-^z
+ \mbox{\boldmath $\sigma_- \cdot$} \hat{\mbox{\boldmath $r$}} \tau_+^z]
Y_1(x_{\pi}),
$$
\begin{equation}
V^{(I=2)}_{P,T} = \overline{g}^{(1)\prime}_{\pi NN} g_{\pi NN}
{m^2_{\pi} \over 8 \pi M}
\mbox{\boldmath $\sigma_- \cdot$}\hat{\mbox{\boldmath $r$}}
[3 \tau_1^z \tau_2^z - \mbox{\boldmath $\tau_1 \cdot \tau_2$}] Y_1(x_{\pi}),
\end{equation}
\noindent where $\mbox{\boldmath $\sigma_+$} = \mbox{\boldmath $\sigma_1$}
 + \mbox{\boldmath $\sigma_2$}$,
 $\mbox{\boldmath $\sigma_-$}
= \mbox{\boldmath $\sigma_1$} - \mbox{\boldmath $\sigma_2$}$ and
similarly for $\tau_+$ and $\tau_-$, $m_{\pi}$ is the pion mass,
 and $M$ the nucleon
mass.  Here $\mbox{\boldmath $r$}
 = \mbox{\boldmath $r_1$}- \mbox{\boldmath $r_2$}$,
 $x_{\pi} = m_{\pi}r$ and $Y_1(x) =
(1 + 1/x) Y_0(x)$ with
$Y_0(x) = e^{-x}/x$.  This result has been given before \cite{one,eight}.
\par
Likewise for the vector mesons, the P,T-violating Lagrangians are
$$
{\cal L}^{(I=0)}_{\rho NN} = i \overline{g}^{(0)
\prime}_{\rho NN} {1 \over 2M}
\overline{N} \sigma_{\mu \nu} \partial_{\nu} \gamma_5 N \tau^a \rho_{\mu}^a,
$$
$$
{\cal L}^{(I=1)}_{\rho NN} = i \overline{g}^{(1)\prime}_{\rho NN} {1 \over 2M}
\overline{N} \sigma_{\mu \nu} \partial_{\nu} \gamma_5 N \rho_{\mu}^z,
$$
\begin{equation}
{\cal L}^{(I=2)}_{\rho NN} = i \overline{g}^{(2)\prime}_{\rho NN} {1 \over 2M}
\overline{N} \sigma_{\mu \nu} \partial_{\nu} \gamma_5 N (3 \tau^z \rho_{\mu}^z
- \tau^a \rho^a_{\mu}),
\end{equation}
\noindent where $\rho^a_{\mu}$ are rho-meson fields and $\sigma_{\mu \nu} =
(\gamma_{\mu} \gamma_{\nu} - \gamma_{\nu} \gamma_{\mu})/2i$.  We are using
the Pauli metric for Dirac matrices as discussed in DeWit and Smith \cite{ten}.
Note that the Lorentz form of these Lagrangians is that of an axial-vector
that is odd
under charge conjugation.  They correspond to second-class axial currents
\cite{eleven} in the
terminology of nuclear $\beta$-decay.  The strong interaction vertex
\begin{equation}
{\cal L}_{\rho NN} = ig_{\rho NN} \overline{N}
(\gamma_{\mu} + i {K_V \over 2M} \sigma_{\mu \nu} \partial_{\nu})
N \tau^a \rho^a_{\mu},
\end{equation}
\noindent where $K_V$ is the ratio of tensor to vector coupling constants. The
corresponding P,T-violating potentials in co-ordinate space are
$$
V^{(I=0)}_{P,T} = - \overline{g}^{(0)\prime}_{\rho NN} g_{\rho NN}
{m^2_{\rho} \over 8 \pi M}
\mbox{\boldmath $\sigma_- \cdot$} \hat{\mbox{\boldmath $r$}}
\mbox{\boldmath $\tau_1 \cdot \tau_2$} Y_1(x_{\rho}),
$$
$$
V^{(I=1)}_{P,T} = \overline{g}^{(1)\prime}_{\rho NN} g_{\rho NN}
{m^2_{\rho} \over 16 \pi M} [\mbox{\boldmath $\sigma_+ \cdot$}
\hat{\mbox{\boldmath $r$}} \tau_-^z
- \mbox{\boldmath $\sigma_- \cdot$}\hat{\mbox{\boldmath $r$}} \tau_+^z]
Y_1(x_{\rho}),
$$
\begin{equation}
V^{(I=2)}_{P,T} = - \overline{g}^{(2)\prime}_{\rho NN} g_{\rho NN}
{m^2_{\rho} \over 8 \pi M} \mbox{\boldmath $\sigma_- \cdot$}
\hat{\mbox{\boldmath $r$}}
(3 \tau_1^z \tau_2^z - \mbox{\boldmath $\tau_1 \cdot \tau_2$}) Y_1(x_{\rho}).
\end{equation}
The overall form of these potentials is the same as for $\pi$-exchange except
 for one
key difference in the isovector potential.  There the relative sign between
 the two
terms is minus for $\rho$-exchange and plus for $\pi$-exchange.  Lastly,
 for
$\omega$-exchange we make substitutions
 $\tau^a \rho^a_{\mu} \rightarrow \omega_{\mu}$
for isoscalar and $\rho^z_{\mu} \rightarrow \tau^z \omega_{\mu}$ for isovector
Lagrangians and obtain the following P,T-violating potentials
$$
V^{(I=0)}_{P,T} = - \overline{g}^{(0)\prime}_{\omega NN} g_{\omega NN}
{m^2_{\omega} \over 8 \pi M} \mbox{\boldmath $\sigma_- \cdot$}
\hat{\mbox{\boldmath $r$}} Y_1(x_{\omega}),
$$
\begin{equation}
V^{(I=1)}_{P,T} = - \overline{g}^{(1)\prime}_{\omega NN} g_{\omega NN}
{m^2_{\omega} \over 16 \pi M} [\mbox{\boldmath $\sigma_+ \cdot$}
\hat{\mbox{\boldmath $r$}} \tau_-^z
+ \mbox{\boldmath $\sigma_- \cdot$}\hat{\mbox{\boldmath $r$}} \tau_+^z]
 Y_1(x_{\omega}).
\end{equation}
\noindent Note in this case there is a relative plus sign between the two
 isovector
terms.  There is no isotensor potential for $\omega$-exchange.
\par
To estimate which components of the P,T-violating interaction are important, we
compute matrix elements of $V_{P,T}$ for closed-shell-plus-one configurations.
This is equivalent to determining effective one-body potentials.  The
computation boils down to an evaluation of two-body matrix elements between the
valence nucleon and one of the nucleons in the closed-shell core summed over
 all the
nucleons in the core.  Let $a$ and $a^\prime$ be two quantum states of the
 valence
nucleon (of the same angular momentum, $j_a$, but opposite parity), and $c$ the
quantum states of the occupied orbits in the core, then for charge-symmetric
cores
$$
<a^\prime m_a |V^{(I)}_{P,T}|a m_a> =
\sum_{c \atop J_1J_2T_1T_2}
{\hat{J}_1 \hat{J}_2 \hat{T}_1 \hat{T}_2 \over (4j_a + 2)}
U \bigl(\mbox{\small $\frac{1}{2} \frac{1}{2}$} I T_2 ;
 T_1 \mbox{\small $\frac{1}{2}$} \bigr)
$$
\begin{equation}
< \mbox{\small  $\frac{1}{2}$} m_a I 0 | \mbox{\small $\frac{1}{2}$} m_a>
<(a^\prime c) J_1 T_1 \| V^{(I)}_{P,T} \| (ac) J_2 T_2>,
\end{equation}
\noindent where $\hat{J} = (2J+1)^{1/2}$ and the $U$-coefficient is a
 recoupling
coefficient of three angular momenta.  Our notation is that of Brink
 and Satchler \cite{twelve}
with the matrix elements reduced in both spin and isospin spaces.
  The two-body matrix
element on the right-hand side of eq. (14) is antisymmetrised; such that
 the sum over
core states includes both the direct and exchange terms.  From selection
 rules, the
U-coefficient is zero when $I=2$, thus there is no contribution to the
 matrix element
from isotensor components of the interaction for charge-symmetric cores.
 Furthermore
matrix elements of \mbox{\boldmath $\sigma_- \tau_1 \cdot \tau_2$} are zero,
 and matrix elements of
$\mbox{\boldmath $\sigma_-$} \tau^z_+$ and $\mbox{\boldmath $\sigma_+$}
\tau_-^z$ are equal when summing over
charge-symmetric cores.
 Thus the $\rho$-exchange interaction gives no contribution in this case.
As noted in \cite{one,eight} a similar result holds for the I=0,2
components of the $\pi$-exchange interaction.
\par
For a charge non-symmetric core $(N \not= Z)$, eq. (14) is generalised to read
$$
<a^\prime m_a|V^{(I)}_{P,T}|a m_a> =
\sum_{{c \atop J_1J_2T_1T_2} \atop m_c M_1 M_2}
{\hat{J}_1 \hat{J}_2 \over (2j_a+1)}
{}~\times
$$
$$
< \mbox{\small $\frac{1}{2}$} m_a \mbox{\small $\frac{1}{2}$} m_c|T_1M_1>
< \mbox{\small $\frac{1}{2}$} m_a \mbox{\small $\frac{1}{2}$} m_c|T_2 M_2>
$$
\begin{equation}
<T_2 M_2 I0|T_1M_1>
<(a^\prime c) J_1 T_1 \| V^{(I)}_{P,T} \| (ac) J_2 T_2>,
\end{equation}
\noindent where Clebsch-Gordan coefficients replace the U-coeff\-icient.
  Here $m_a$ is
the $z$-compon\-ent of the valence nucleon's isospin quantum number and is
 $-1/2$ for a
neutron and $+1/2$ for a proton.  Now terms that gave zero contribution for
charge-symmetric cores give a finite but generally small contribution
 in heavy nuclei.
Some sample calculations are given in Table \ref{contrib}
for $j_a = 1/2$, $p$- and $s$- states.
\par
Our calculations are made with harmonic oscillator wavefunctions.
  However there is
one drawback to this choice; the eigenfunctions of the harmonic oscillator
 potential
are not the eigenfunctions of the nucleon-nucleon interaction.
  This is particularly
important in the relative coordinate, $r$, where the nucleon-nucleon
 interaction is
known to have a strong short-range repulsion, which makes the relative
wavefunction go
rapidly to zero as $r \rightarrow 0$, more rapidly than given by uncorrelated
oscillator functions.  Thus to incorporate this piece of many-body physics
 in a simple
way it is quite common to modify two-body operators by multiplying them by a
short-range correlation function.  Thus we write
\begin{equation}
\hat{V}^{(I)}_{P,T}(r) = V^{(I)}_{P,T}(r) \hat{g}(r),
\end{equation}
\noindent where $\hat{g}(r)$ is some function that tends to zero as
 $r \rightarrow 0$
and tends to unity for large $r$.  The precise choice of $\hat{g}(r)$
 becomes part of
the model dependence at short distances.  This choice is not critical if
 the operator
is of long-range, such as the pion-exchange interactions, but is much
 more critical
for the shorter-ranged, heavy-meson exchange interactions.
 Our results in Table \ref{contrib}
adopt the choice made by Adelberger and Haxton \cite{thirteen}, who
 in their work on parity violation
in nuclei used $\hat{g}(r) = [1 - exp(-ar^2)(1-br^2)]^2$
 with $a = 1.1 fm^{-2}$
and $b = 0.68 fm^{-2}$ obtained from the work of Miller and Spencer
\cite{fourteen}.  It is
evident from Table \ref{contrib} that the short-range correlation function
significantly reduces
the contribution from heavy mesons. Thus, providing all the unknown coupling
constants $\overline{g}^{(I)\prime}_{MNN}$ are of comparable magnitude,
 it can be
asserted that the isovector pion-exchange component will dominate
 P,T-violation in
nuclei.
\par
An estimate of the coupling constants
has  recently been provided by
 Gudkov, He and MacKellar \cite{fifteen}.
These authors point out that, in general, a first principles calculation of
P,T-violation would be very difficult. However, beginning with an effective
 low-energy
P,T-violating Lagrangian at the quark level for particular models of
CP violation,
they have estimated the P,T-violating coupling constants,
$\overline{g}^{(I)\prime}_{MNN}$, of the nucleon-nucleon interaction using a
factorization approximation and the vector dominance hypothesis.  Their main
conclusion is that for all types of models of CP violation in the
 one-meson-exchange
approximation the contributions to the P,T-violating nucleon-nucleon
 interaction from
pseudoscalar mesons are larger than the contributions from vector mesons
 by about one
order of magnitude.

\section{THE RATIO $\kappa^{(1)}$} \label{sec:kappa}
We now return to the evaluation of the ratio of P,T-violation to
 P-violation matrix
elements of eq. (6).  We assume that P-violation principally comes from the
 isoscalar
$\rho$-exchange potential \cite{six,thirteen}
\begin{equation}
V_{\rho} = -i g^{(0)\prime}_{\rho NN} (1 + K_V)
{m^2_{\rho} \over 4\pi M}
i \mbox{\boldmath $\sigma_1 \times \sigma_2 \cdot$} \hat{\mbox{\boldmath $r$}}
\;
\mbox{\boldmath $\tau_1 \cdot \tau_2$}
Y_1(x_{\rho}),
\end{equation}
where we have neglected a smaller non-local term.  The weak coupling constant
$g^{(0)\prime}_{\rho NN}$ is the same as $h^0_{\rho}$ in the notation of
 DDH \cite{six}.  The
ratio therefore becomes
\begin{eqnarray}
\kappa^{(1)} &= &{1 \over 4} {g_{\pi NN} \over g_{\rho NN} (1+K_V)}
{m^2_{\pi} \over m^2_{\rho}} \times
\nonumber
\\
&&{\langle (\mbox{\boldmath $\sigma_+ \cdot$}\hat{\mbox{\boldmath $r$}}
\mbox{\boldmath $\tau_- +
\sigma_- \cdot$}\hat{\mbox{\boldmath $r$}} \;
\mbox{\boldmath $\tau_+$})Y_1(x_{\pi})\rangle
\over
\langle i \mbox{\boldmath $\sigma_1 \times \sigma_2 \cdot$}
\hat{\mbox{\boldmath $r$}}
\;  \mbox{\boldmath $\tau_1 \cdot \tau_2$} Y_1(x_{\rho}) \rangle}
{\overline{\omega}_{\pi} \over {\omega}_{\rho}},
\end{eqnarray}
\smallskip

\noindent where $\overline{\omega}_{\pi}$, ${\omega}_{\rho}$
 are reduction factors accounting for short-range
correlations.
\par
Our exact numerical results for the  ratio of the P,T-violating to P-violating
matrix elements, (incorporating the reduction
factors $\overline{\omega}_{\pi}$ and
 ${\omega}_{\rho}$
explicitly as decribed in eq. (16) ), are summarized
in Table \ref{kappa}. There we give some sample neutron matrix elements
in closed-shell-plus-one
configurations for $j_a = 1/2$, $p$- and $s$-states.
  Significantly, the $\kappa^{(1)}$
 values display
very little mass dependence, and range from 2 to 5.

To understand our results summarized in Tables I and II
let us consider the simple case
a neutron in a
closed-shell-plus-one configuration for charge symmetric cores, eq. (14).
  Then, for
the $\pi$-exchange term, the exchange component of the two-body matrix
 elements is
identically zero, while for the $\rho$-exchange term the direct component
 is zero.
The spin and isospin sums are trivially evaluated:
 $<\mbox{\boldmath $\sigma_+$} \tau_-> =
\linebreak
<\mbox{\boldmath $\sigma_-$} \tau_+> = 12$ and
$<i \mbox{\boldmath $\sigma_1\times\sigma_2  \tau_1\cdot\tau_2$}>
 = -18$
and the angle integral $<\hat{\mbox{\boldmath $r$}}>$
 cancels in the ratio to give
\smallskip

\begin{equation}
\kappa^{(1)} = - {1 \over 3}
{g_{\pi NN} \over g_{\rho NN} (1 + K_V)}
{m^2_{\pi} \over m^2_{\rho}}
{<Y_1(x_{\pi})> \over <Y_1(x_{\rho})>}
{\overline{\omega}_{\pi} \over {\omega}_{\rho}}.
\end{equation}
\smallskip
It remains to evaluate the radial integrals, $<Y_1(x_{\pi})>$.
 Consider the integral evaluated between $0s$ and $0p$ oscillator
functions of length parameter, $b$.  The exact value of the integral is
\begin{equation}
<Y_1(x)> = ({2 \over 3})^{1/2} [z e^{z^2} erfc(z) + {1 \over \pi^{1/2}}
({1 \over 2z^2} - 1)]
\end{equation}
$$
\Rightarrow({2 \over 3})^{1/2} {3 \over 4\pi^{1/2}} {1 \over z^4}
\qquad \qquad \hbox {for} \qquad z >> 1,
$$

\noindent where $z = mb/ \sqrt 2$ and $m$ the meson mass.
Thus, in the limit of
large meson mass the radial integrals, $<Y_1(x_{\pi})>$ scales
as $1/m^4$.  The short-range correlation function
cuts down matrix elements of $\rho$-meson range by roughly
a factor of three more than those of $\pi$-meson range,
$\overline{\omega}_{\pi}/{\omega}_{\rho}\simeq 3$
\par
The approximate scaling of the radial integrals with $1/m^4$
 is largely responsible
for the large difference in the magnitudes of the $\pi$- and vector meson
P,T-violating
 matrix elements.
We note, however, that for pion range, $m_{\pi} =
0.7 fm^{-1}$, the inequality $z >> 1$ is not satisfied, not even in
 heavy nuclei since
$b$ only varies gently with nuclear mass as $A^{1/6}$.  This results in
the scaling
 approximation, when taken at face value,
leading to an overestimate in the ratio $\kappa^{(1)}$ by a
 factor of three.
More detailed calculations reduce $\kappa^{(1)}$ even further, to give
the values listed in the Table II.
\par
 In addition to  examining\cite{nine} the effective one-body
P,T-violating potential,
Griffiths and Vogel \cite{four}
has also calculated $\kappa^{(1)}$  for specific rare-earth nuclei,
 using the quasiparticle random phase approximation.
They find $\kappa^{(1)}$ values in the range 1 to 6,
which is similar to the present range of
2 to 5.  Since $\kappa^{(1)}$ seems
 to be determined to
within quite a small range independent of nuclear structure considerations,
 then a
degree of proportionality between the matrix elements of the
P,T- and P-violating
potentials seems to be established.

\section{P,T-Violation in neutron transmission} \label{sec:neutrons}
 There is much interest in the possibility of probing
P,T-violation
in the transmission of polarized
neutrons through polarized nuclear targets \cite{two}.
Neutron transmission measurements on unpolarized targets have proved
to be powerful probes of P-violation. There the scattering cross sections of
low-energy neutrons from nuclei at $p$-wave resonances
exhibit very large parity violating
longitudinal asymmetries.  The P-violating asymmetries,
 which are defined as the
fractional difference of the resonance cross section for neutrons polarized
parallel and antiparallel to their momentum, $P_P =
\sigma^+ - \sigma^-/ \sigma^+ +\sigma^-$, arise from
$s_{1/2}$ admixtures in the $p_{1/2}$ resonances, and can be as
large as 10\%. In the case of polarized
targets the asymmetry  $P_{P,T} =
\sigma^{\uparrow} - \sigma^{\downarrow} / \sigma^{\uparrow}
+\sigma^{\downarrow}$,
implies simultaneous
P- and T-violation  through a term in the scattering amplitude proportional
to $<\mbox{\boldmath$\sigma$}>\cdot (\mbox{\boldmath$k$} \times
 <\mbox{\boldmath$J$}> )$.
Here
$\sigma^{\uparrow}$ and $\sigma^{\downarrow}$ are the total scattering
cross-sections for neutrons polarized parallel and antiparallel to
$(\mbox{\boldmath$k$} \times
 <\mbox{\boldmath$J$}> )$, and $\mbox{\boldmath$\sigma$}$ and
$\mbox{\boldmath$k$}$ are the neutron spin and momentum and
 $\mbox{\boldmath$J$}$ is the target spin.

In analogy with eq. (5), the ratio of the P,T-violating to P-violating
asymmetries in the neutron measurements
can be related to the ratio of the matrix elements of the
corresponding weak NN potentials. In a two-state mixing approximation the ratio
is given by,
\begin{equation}
\lambda  =  \frac{P_{P,T}}{P_P} \simeq
 -i  {<\psi_s|V^{P,T}|\psi_p> \over
<\psi_s|V^P|\psi_p>},
\end{equation}
where $\psi_s$ and $\psi_p$ are compound nuclear $s$- and $p$-wave resonance
wave functions. To first order the ratio of the weak matrix elements
 between compound resonances can be approximated by the ratio of
single-particle matrix elements, so that \cite{one},
\begin{equation}
\lambda^{(I)} \approx
 \kappa^{(I)} {\overline{g}^{(I)\prime}_{\pi NN} \over g^{(0)\prime}_{\rho
NN}}.
\end{equation}
Herczeg has reviewed
the experimental searches of electric dipole moments in atoms, molecules and
the neutron, and barring cancellations between the different isospin
amplitudes,  deduces the following upper limits for the P,T-violating
$\pi$NN constants
$$
\mid \overline{g}^{(0)\prime}_{\pi NN}\mid \buildrel < \over \sim
 1.4\times 10^{-11},
$$
\begin{equation}
\mid \overline{g}^{(1)\prime}_{\pi NN}\mid
 \buildrel < \over \sim 10^{-10},
\end{equation}
$$
  \mid \overline{g}^{(2)\prime}_{\pi NN}\mid \buildrel < \over \sim
 1.4\times 10^{-11} .
$$
To get a rough estimate of the accuracy
needed in the neutron transmission experiments
to improve on these limits we consider $^{139}$La,
where a very large P-violating asymmetry
 ($P_{P} \sim 10^{-1}$) has been observed.
Combining Herczeg's limits with the values of $\kappa^{(0)}, \kappa^{(1)}$,
 and $\kappa^{(2)}$ from Tables I
and II   for A=138, and taking $g^{(0)\prime}_{\rho NN}= -11.4\times 10^{-7}$
gives,
\begin{equation}
\mid \lambda^{(0)} \mid \buildrel < \over \sim  1.6\times 10^{-5},
\end{equation}
\begin{equation}
\mid \lambda^{(1)} \mid \buildrel < \over \sim  5\times 10^{-4},
\end{equation}
\begin{equation}
\mid \lambda^{(2)} \mid \buildrel < \over \sim 2.5\times 10^{-5}.
\end{equation}
The enhanced sensitivity to the isovector coupling arises
because the $I=1$ P,T-violating interaction is probed by all A nucleons,
whereas the isoscalar and isotensor interactions are only probed by
the excess neutrons \cite{one}.
This situation is the reverse of that in the neutron electric
dipole experiments, where sensitivity to the isovector term is suppressed.
 A
$P_{P,T}$/P$_P$ $\sim$ 10$^{-4}$ would then provide
much tighter constraints on the isovector P,T-odd $\pi$NN coupling constant.

\section{CONCLUSIONS} \label{sec:conc}

We have derived the PT-violating NN potentials for $\rho$- and
$\omega$-exchange
and demonstrated quantitatively what has only been speculated qualitatively
before, namely that $\rho$ and $\omega$
exchange  give a negligible contribution
to time-reversal violation compared
to $\pi$ exchange, assuming comparable coupling constants.
Under this assumption the heavy-meson matrix elements are about two orders
of magnitude smaller than the $I=1$ $\pi$-exchange matrix elements.
 This is in sharp contrast to parity violation,
where heavy-meson exchange is comparable to pion exchange.

The supression of the vector-meson contributions to P,T-violation
arises from a combination of factors.  The first of these applies to
 $\rho$-exchange where the form of the P,T-violating potentials
 are the same as for $\pi$-exchange
except for a crucial sign difference between the two terms
contributing to the  $I=1$
potential. For $N=Z$ cores these two terms are equal in magnitude. They add
constructively for $\pi$-exchange but exactly
cancel for $\rho$-exchange. Moving to heavier nuclei, where $N \neq Z$,
breaks this cancellation to a small though not significant degree.
The $I=0$ and $I=2$ $\rho$-exchange
 potentials have identical forms to the corresponding
$\pi$-exchange potentials. Thus, as in the case of the $I=0$ and $I=2$ pion
terms, they give zero contribution to P,T-violation in $N=Z$ systems, and
contribute through the excess neutrons in heavy nuclei. We note,
as can be seen from Table I, that
the $I=0$ and $I=1$ $\omega$-exchange terms are not
zero for charge-symmetric systems, owing to the different nature of the
 $\omega$-exchange potentials, eqs. (13).

The second source of suppression
 applies equally to $\rho$- and $\omega$-exchange
and is due to the {\em approximate} scaling of the radial matrix of
$Y_1(x)$  with $1/m^4$, so that the matrix elements
of $V_{P,T}$ scale approximately as $1/m^2$.  This reduces
the $\rho$- and $\omega$-exchange
matrix elements by about a factor of 25 relative to those for
the pion.
 Finally, the short-range
correlations function
reduces the radial matrix elements for the heavy-mesons by
an additional factor of about three.

P,T-violation in both $\gamma$-decay and neutron transmission experiments is
always accompanied by P-violation alone, and the sensitivity of these searches
is determined by the ratio of the P,T-violating to P-violating matrix elements.
We have calculated this ratio, $\kappa^{(1)}$, for closed-shell-plus-one
configurations and find it to be approximately independent of
 the nucleus under study. We find the
 calculated values of $\kappa^{(1)}$ to lie in the range 2 to 5. This range is
similar to the range 1 to 6 found by Griffiths and Vogel \cite{four} in
calculations for
rare earth nuclei using the quasiparticle random phase approximation.
This  suggests that $\kappa^{(1)}$ is determined within quite
a small range independent of detailed nuclear structure considerations.
 Thus, the degree of proportionality
between time-reversal-violation and parity-violation matrix elements seems
to have been established.

\bigskip

The authors wish to thank Peter Herczeg for correspondence on the weak
 $\rho$-meson
Lagrangians, and for valuable suggestions concerning the manuscript.

\bigskip

%\begin{thebibliography}{99}

%\end{thebibliography}
\newpage
\onecolumn
\squeezetable
\mediumtext
\begin{table}
%\noindent {\bf Table 1:} Contribution from each component of the two-body
\caption[Table I]{Contribution from each component of the two-body
$V_{P,T}$ interaction to the P,T-violating matrix element in
closed-shell-plus-one configuration in units of $eV \times 10^{-4}$ for six
choices of the closed-shell core. Matrix elements were calculated with harmonic
oscillator wave functions with $\hbar\omega = 45A^{-1/3}- 25A^{-2/3}$ MeV. The
Miller-Spencer [14] short-range correlation function was used. The weak
interaction coupling constants, $\overline{g}^{(I)\prime}_{MNN}$, were set at
$ 1\times 10^{-11}$.}
%$ 1\times 10^{-11}$.\\

\begin{tabular}{l r r r r r r}
& & & & & & \\
 & $^{16}$O & $^{40}$Ca
& $^{90}$Zr & $^{138}$Ba &  $^{208}$Pb & $^{232}$Th \\
 & N=8 & N=20
& N=50 & N=82
& N=126 & N=142 \\
 & Z=8 & Z=20
& Z=40 & Z=56
& Z=82 & Z=90 \\
 & & & & & & \\ \hline
 & & & & & & \\
 & $\underline{0p-0s}$ & $\underline{1p-1s}$
& $\underline{2p-2s}$ & $\underline{2p-2s}$
& $\underline{3p-3s}$ & $\underline{3p-3s}$ \\
 & & & & & & \\
 $\pi$(I=0) & - & - & $-0.059$
& $-0.181$ & $-0.126$ & $-0.162$ \\
 $\pi$(I=1) & 1.084 & 0.875
& 0.708 & 0.779 & 0.608
& 0.633 \\
 $\pi$(I=2) & - & -
& 0.000 & $-0.275$ & $-0.190$ & $-0.179$ \\
 & & & & & & \\
 $\rho$(I=0) & - & -
& 0.000 & 0.002 & 0.001
& 0.001 \\
 $\rho$(I=1) & - & -
& $-0.000$ & $-0.001$ & $-0.000$
& $-0.001$ \\
 $\rho$(I=2) & - & -
& 0.000 & 0.001 & 0.000
& 0.000 \\
 & & & & & & \\
 $\omega$(I=0) & 0.020 & 0.012
& 0.007 & 0.008 & 0.006
& 0.006 \\
 $\omega$(I=1) & $-0.020$ & $-0.012$
& $-0.007$ & $-0.008$ & $-0.006$ & $-0.006$ \\
& & & & & & \\
 & $\underline{0p-1s}$ & $\underline{1p-2s}$
& $\underline{2p-3s}$ & $\underline{2p-3s}$
& $\underline{3p-4s}$ & $\underline{3p-4s}$ \\
 & & & & & & \\
$\pi$(I=0) & - & - & 0.054
& 0.122 & 0.097 & 0.125 \\
 $\pi$(I=1) & $-0.400$ & $-0.378$
& $-0.388$ & $-0.465$ & $-0.376$
& $-0.409$ \\
 $\pi$(I=2) & - & -
& 0.000 & 0.181 & 0.146
& 0.138 \\
 & & & & & & \\
 $\rho$(I=0) & - & -
& $-0.000$ & $-0.001$ & $-0.001$
& $-0.001$ \\
 $\rho$(I=1) & - & -
& 0.000 & 0.000 & 0.000
& 0.001 \\
 $\rho$(I=2) & - & -
& 0.000 & $-0.000$ & $-0.001$ & $-0.001$ \\
 & & & & & & \\
 $\omega$(I=0) & $-0.008$ & $-0.003$
& $-0.004$ & $-0.005$ & $-0.003$
& $-0.004$ \\
 $\omega$(I=1) & 0.008 & 0.003
& 0.004 & 0.005 & 0.003
& 0.004 \\
 & & & & & & \\
\end{tabular}
\label{contrib}
\end{table}

%\vfil\eject
\onecolumn
\squeezetable
\mediumtext
\begin{table}
\caption[Table II]{Isovector $\pi$-exchange, $V_{P,T}$, and isoscalar
%\noindent {\bf Table 2:}  Isovector $\pi$-exchange, $V_{P,T}$, and isoscalar
$\rho$-exchange, $V_P$, matrix elements evaluated for a closed-shell-plus-one
configuration for six choices of the closed-shell core.  The weak interaction
coupling
constants are $\overline{g}^{(1)\prime}_{\pi NN} = 1.0 \times 10^{-11}$ and
$g^{(0)\prime}_{\rho NN} = -11.4 \times 10^{-7}$.  Matrix elements were
calculated
with harmonic oscillator wavefunctions with $\hbar \omega
= 45A^{-1/3} - 25A^{-2/3}$
MeV.  The Miller-Spencer [14] short-range correlation function was used.
The ratio,
$\kappa^{(1)}$, is defined in eq. (6).}
%$\kappa^{(1)}$, is defined in eq. (6).\\

\begin{tabular}{ c r r r r r r}
 & & & & & & \\
 & $^{16}$O & $^{40}$Ca
& $^{90}$Zr & $^{138}$Ba
& $^{208}$Pb & $^{232}$Th \\
 & N=8 & N=20 & N=50 & N=82
& N=126 & N=142 \\
 & Z=8 & Z=20
& Z=40 & Z=56 & Z=82 & Z=90 \\
 & & & & & & \\ \hline
 & & & & & & \\
 & $\underline{0p-0s}$ & $\underline{1p-1s}$
& $\underline{2p-2s}$ & $\underline{2p-2s}$
& $\underline{3p-3s}$ & $\underline{3p-3s}$ \\
 & & & & & & \\
 & & & & & & \\
 $<V_{P,T}>$ in eV $\times 10^{-4}$
& 1.084 & 0.875
& 0.708 & 0.779
& 0.608 & 0.633 \\
 $i<V_P>$ in eV
& 1.513 & 1.550
& 1.535 & 1.576 & 1.581 & 1.600 \\
 & & & & & & \\
 $\kappa^{(1)}$
& $-8.2$ & $-6.4$
& $-5.3$ & $-5.6$
& $-4.4$ & $-4.5$ \\
 & & & & & & \\
 & & & & & &  \\
 & $\underline{0p-1s}$ & $\underline{1p-2s}$
& $\underline{2p-3s}$ & $\underline{2p-3s}$
& $\underline{3p-4s}$ & $\underline{3p-4s}$ \\
& & & & & & \\
& & & & & & \\
 $<V_{P,T}>$ in eV $\times 10^{-4}$
& $-0.400$ & $-0.378$
& $-0.388$ & $-0.465$
& $-0.376$ & $-0.409$ \\
 $i<V_P>$ in eV
& 1.294 & 1.435
& 1.441 & 1.485
& 1.508 & 1.527 \\
 & & & & & & \\
 $\kappa^{(1)}$
& 3.5 & 3.0
& 3.1 & 3.6
& 2.8 & 3.0 \\
 & & & & & & \\
\end{tabular}
\label{kappa}
\end{table}
\end{document}